\documentclass[aps,rmp,reprint,showkeys,groupedaddress]{revtex4-2}
\usepackage{graphicx}
\usepackage{bm}
\usepackage{xcolor}

\begin{document}

\newcommand{\be}{\begin{equation}}
\newcommand{\ee}[1]{\label{#1}\end{equation}}
\newcommand{\bem}{\begin{eqnarray}}
\newcommand{\eem}[1]{\label{#1}\end{eqnarray}}
\newcommand{\eq}[1]{Eq.~(\ref{#1})}
\newcommand{\Eq}[1]{Equation~(\ref{#1})}
\newcommand{\rc}[1]{\textcolor{red}{#1}}


\title{ Spin superfluidity }

\author{E.  B. Sonin}
 \affiliation{Racah Institute of Physics, Hebrew University of
Jerusalem, Givat Ram, Jerusalem 9190401, Israel}

\date{\today}

\begin{abstract}

The phenomenon of superfluidity (superconductivity) is a possibility of transport of mass (charge) on macroscopical distances without essential dissipation. In magnetically ordered media with easy-plane topology of the order parameter space the superfluid transport of spin is also possible despite the absence  of the strict conservation law for spin. The article addresses three key issues in the theory of spin superfluidity: topology of the order parameter space, Landau criterion for superfluidity, and decay of superfluid currents via phase slip events, in which magnetic vortices cross current streamlines. Experiments on detection of spin superfluidity are also surveyed. 

\end{abstract}


\maketitle

\onecolumngrid
Key Points/Objective
\begin{itemize}
\item Spin superfluidity is the possibility to transport spin with essentially suppressed dissipation on long distances.

\item Spin superfluidity is possible if the magnetic order parameter space has the topology of a circumference.

\item The necessary topology is provided by easy-plane anisotropy in ferromagnets or by magnetic field in antiferromagnets.

\item Metastability of spin superfluid current states is restricted by the Landau criterion.

\item  Decay of spin superfluid currents is realized via phase slips, in which magnetic vortices cross current streamlines.
\end{itemize}

\vspace{5mm}

\twocolumngrid

\tableofcontents

\renewcommand{\thesection}{\arabic{section}}


\section{Introduction} \label{Intr}

 The phenomenon of superfluidity  (superconductivity in the case of charged fluids) is known  more than a hundred years. Its analog for spin (spin superfluidity) 
 occupies minds of condensed matter physicists from 70s of the last century.  The term {\em superfluidity} is used in the literature to cover a broad range of phenomena in superfluid $^4$He and $^3$He, Bose-Einstein condensates of cold atoms, and solids.  In this article  superfluidity  means 
 only the possibility to transport a physical quantity (mass, charge, spin, ...) without dissipation (more accurately, with essentially suppressed dissipation). This corresponds to the original hundred-years old meaning of the  term from the times of Kamerlingh Onnes and Kapitza. 

  The superfluidity is conditioned by the special  topology of the order parameter space (vacuum manifold).  Namely, this topology is that of a circumference in a plane.  The angle of rotation around this circumference is the order parameter phase describing all degenerate ground states.    In superfluids the phase is  the phase of the macroscopic wave function.
  In  magnetically ordered systems (ferro- and antiferromagnets) the necessary topology is provided by  easy-plane magnetic anisotropy, and the phase is  the angle of rotation around the axis (further the axis $z$) normal to the easy plane. Currents of mass (charge)  or spin are proportional to phase gradients and  are called supercurrents. 
    
 In early discussions of the spin supercurrent  it was considered as a counterflow of superfluid currents of particles with different spins in superfluid $^3$He \cite{Vuorio74}, i.e., spin was transported by itinerant spin carriers. Later it was demonstrated that spin superfluidity is a universal phenomenon, which does not require mobile spin carriers and is possible in magnetic insulators  \cite{ES-78b,ES-82}. It can be described within the framework of  the  standard Landau--Lifshitz--Gilbert (LLG) theory. However, the publications conditioning spin superfluid transport by the presence of mobile carriers of spin continued to appear in the literature \cite{Bun,Niu}.  According to \citet{Niu}, it is  a critical  flaw of spin-current definition if it predicts spin currents in insulators. 

Strictly speaking the analogy of spin superfluidity with superfluids is complete only if there is invariance with respect  to any rotation in the spin space around the axis $z$. Then according to Noether's theorem the spin component along the axis $z$ is conserved. But there is always some magnetic anisotropy, which breaks the rotational  invariance. Correspondingly, there is no strict conservation law for spin, while in superfluids the gauge invariance  is exact, and  the  conservation law of mass (charge) is also exact.

In the past  there were arguments about whether  superfluidity  is possible in the absence of the conservation law. This dispute started at discussion of superfluidity of electron-hole pairs or excitons. The number of electron-hole pairs can vary due to interband transitions, and  the degeneracy with respect to the phase of the pair condensate is lifted.  \citet{GuKe} called this  effect ``fixation of phase''.  They demonstrated that spatially \emph{homogeneous} stationary current states are impossible, and concluded that there is no analogy with superfluidity.  However,  later it was demonstrated that phase fixation does not rule out the existence of weakly \emph{inhomogeneous} stationary current states analogous to superfluid current states \cite{KuSh,ES-77,LY}. This analysis was extended on spin superfluidity \cite{ES-78a,ES-78b,ES-82}.
In magnetism violation of the spin conservation law  usually is rather weak because it is related with relativistically small (inversely proportional to the speed of light) processes of spin-orbit interaction. In fact, the LLG theory itself is based on the assumption of weak spin-orbit interaction \cite{LLstPh2}.

Above we discussed supercurrents generated in the equilibrium (ground state) of the magnetically ordered medium with easy plane topology. But  in magnetically ordered system this topology  is possible also in non-equilibrium coherent precession states, when spin pumping supports spin precession with fixed spin component along the magnetic field. Such non-equilibrium coherent precession states, which are called nowadays magnon BEC, were experimentally investigated in the $B$ phase of superfluid $^3$He and in yttrium-iron-garnet (YIG) films \cite{Bun,Dem6}. 

Spin superfluid transport  is possible as long as the spin phase gradient does not exceeds the critical value determined by the Landau criterion.   The Landau criterion checks stability of supercurrent states with respect to elementary excitations  of all collective modes. The Landau criterion determines a  threshold for the current state instability, but it tells nothing about how the instability develops. The decay of the supercurrent is possible only via phase slips. In a phase slip event a magnetic  vortex crosses current streamlines decreasing the phase difference along streamlines. Below the critical value of supercurrent phase slips are suppressed by energetic barriers. The critical value of the supercurrent at which barriers vanish is of the same order as that estimated from the Landau criterion. This leads to a conclusion that the instability predicted by the Landau criterion is a precursor of the avalanche of phase slips not suppressed by any activation barrier.

The superfluid spin transport on macroscopical distances is possible only if the  spin phase performs a large number of full $2\pi$ rotations along current streamlines (large winding number), and phase slips are suppressed by energetic barriers. On the other hand, small phase variations less than  $2\pi$ are ubiquitous in magnetism. They emerge in any spin wave, any domain wall, or due to disorder. Their existence is confirmed by numerous experiments in the past. Spin currents generated by these small phase differences transport spin only on small distances and oscillate in space or time, or both.  Their existence is not a manifestation of spin superfluidity. 

In last decades the interest to spin superfluidity was revived \cite{Adv,BunV,Tserk,Halp,Pokr,Mac,Son17,Hoefer,TserkKlaui,BratAF,Son19,Nowak} in connection with the  emergence of spintronics.  The present article reviews the three essentials of the spin superfluidity concept: topology, Landau criterion, and phase slips. The article focuses on the qualitative analysis avoiding details of calculations, which can be found in original papers. After the theoretical analysis the experiments searching for   spin superfluidity are shortly discussed. 

\section{Concept of superfluidity}\label{MS}

\begin{figure}[b]
\includegraphics[width=.45\textwidth]{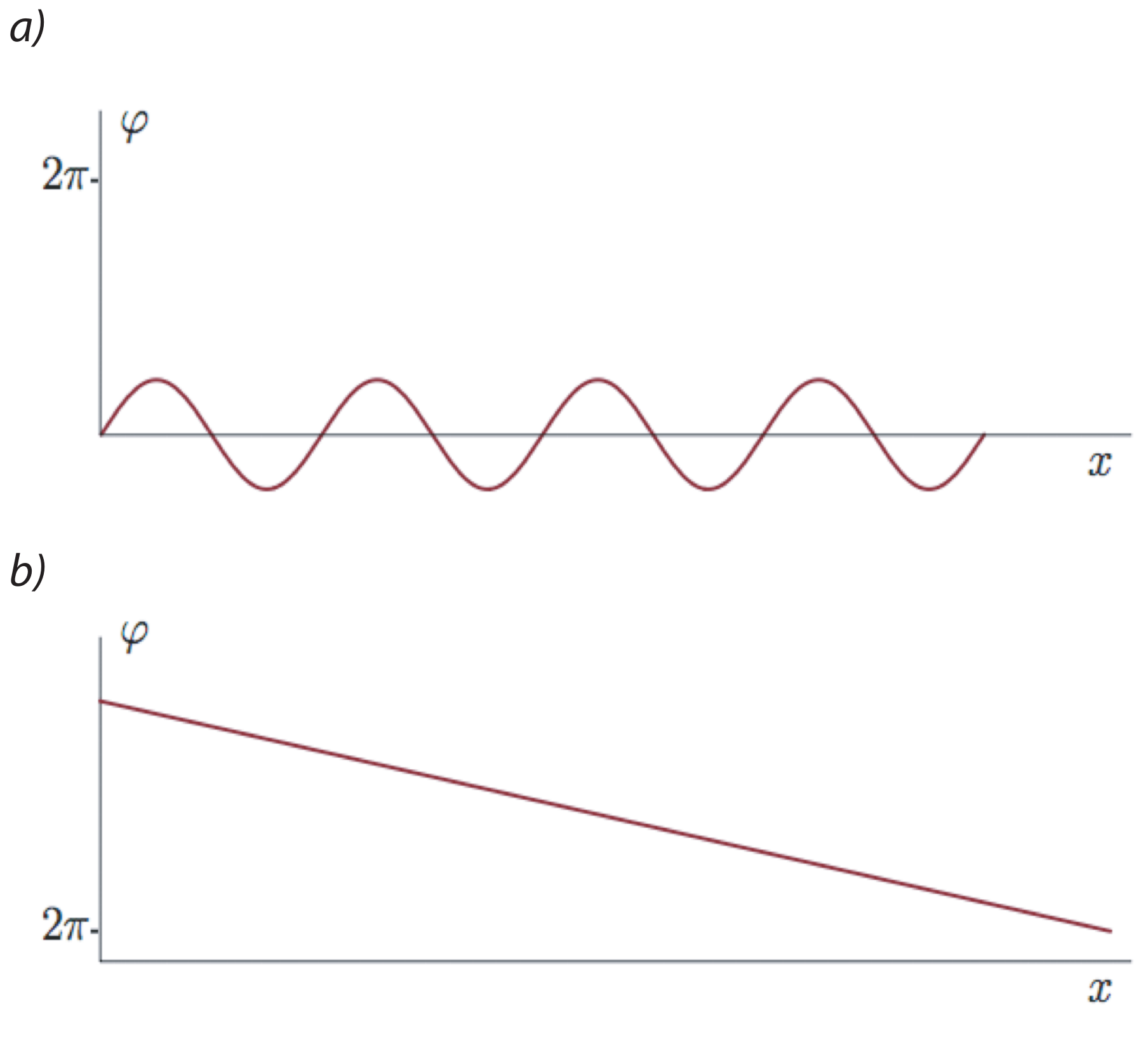}
\caption[]{ Phase (in-plane rotation angle)  variation in space at the presence of mass (spin) currents [From \citet{Adv}]. (a) Oscillating currents in a sound (spin) wave. (b) Stationary mass (spin) supercurrent. }
\label{fig1}
\end{figure}

Let us remind the concept of superfluidity for the transport of mass (charge).
In superfluid hydrodynamics there are the Hamilton equations for the pair of the canonically conjugate variables ``phase - particle density'':
 \begin{eqnarray}
\hbar {d\varphi \over dt}=-{\delta {\cal H}\over \delta n},~~
{dn \over dt}= {\delta{\cal H}\over \hbar \delta \varphi}.
     \label{IdHam} \end{eqnarray}
Here  
\begin{equation}
{\delta {\cal H}\over \delta n}={\partial  {\cal H}\over \partial n} -\bm \nabla \cdot {\partial  {\cal H}\over \partial \bm \nabla n},~~{ \delta {\cal H}\over \delta \varphi}={\partial  {\cal H}\over \partial \varphi} -\bm \nabla \cdot {\partial  {\cal H}\over \partial \bm \nabla \varphi} 
         \end{equation}
are functional derivatives of the Hamiltonian 
\be
{\cal H} ={\hbar^2n \over 2m}\nabla \varphi^2+ E_0(n),
    \ee{}
$\varphi$ is the phase of the wave function describing the Bose-Einstein condensate (BEC) in Bose liquids or the Cooper pair condensate in Fermi liquids, and
$E_0(n)$ is the energy of the superfluid at rest, which depends only on the particle density $n$. Taking into account  the gauge invariance [$U(1)$ symmetry] $\partial {\cal H} / \partial \varphi=0$,  the Hamilton equations are reduced to  the equations of hydrodynamics for an ideal liquid:
  \begin{eqnarray}
m{d\bm v \over dt}= -\bm \nabla \mu ,
\label{Eul}           \end{eqnarray}
 \begin{eqnarray}
{dn \over dt}=-\bm \nabla \cdot \bm j.
     \label{IdLiq} \end{eqnarray}
In these expressions 
\be
\mu= {\partial E_0\over \partial n}+{\hbar^2 \over 2m}\nabla \varphi^2
    \ee{}
is the chemical potential, and
\begin{equation}
 \bm j=n\bm v ={\partial  {\cal H}\over \hbar \partial \bm \nabla \varphi} 
 \end{equation} 
 is the particle current. We consider the zero-temperature limit, when the superfluid velocity coincides with the center-of-mass velocity of the whole liquid:
 \be
 \bm v= {\hbar \over m} \bm \nabla \varphi.
          \ee{}
The continuity equation \eq{IdLiq}  satisfies the conservation law of mass (charge), which follows from the gauge invariance.

A collective mode of the ideal liquid is a plane sound wave $\propto e^{i\bm k\cdot \bm r-i\omega t} $ with the wave vector $\bm k$, the frequency $\omega$, and  the linear spectrum $\omega =u_s k$. The sound velocity is 
\be
u_s =\sqrt{{n\over m}{\partial^2 E_0 \over \partial n^2}}.
     \ee{}
In the sound wave the phase varies in space, i.e., the wave is accompanied by mass currents [Fig.~\ref{fig1}(a)]. An amplitude of the phase variation is small, and currents transport mass on distances of the order of the wavelength. A real superfluid transport on macroscopic distances is possible in current states, which are stationary solutions of the hydrodynamic equations with finite constant currents, i.e., with constant nonzero phase gradients. In the current state the phase rotates through a large number of full 2$\pi$-rotations along streamlines of the current [Fig.~\ref{fig1}(b)]. These are supercurrents or persistent currents.  

The crucial point of the superfluidity theory is why the supercurrent in Fig.~\ref{fig1}(b) is  a metastable  state and does not decay a long time. 
The first explanation of the supercurrent metastability was the well known Landau criterion \cite{Lan}. According to this criterion, the current state is stable as far as \emph{any quasiparticle}  in a moving liquid has a positive energy  in the laboratory frame and therefore its creation requires an energy input. Let us suppose that elementary quasiparticles of  the liquid at rest have an energy $\varepsilon(\bm p)=\hbar\omega (\bm k)$,  where $\bm p=\hbar \bm k$ is the quasiparticle momentum. If the liquid moves with the velocity $\bm v$ the quasiparticle energy in the laboratory frame becomes $\tilde\varepsilon(\bm p)=\varepsilon(\bm p)+\bm p\cdot \bm v$. This is the Doppler effect in the Galilean invariant fluid.
The current state is stable if the energy $\tilde\varepsilon$ is never negative. This condition is the Landau criterion:  
\begin{equation}
v < v_L =\mbox{min}{\varepsilon(\bm p)\over p}=\mbox{min}{\omega(\bm k)\over k}.
  \label{LC}       \end{equation}
If quasiparticles are phonons (quanta of sound waves) the Landau critical velocity $v_L$ is the sound velocity $u_s$. In superfluid $^4$He the Landau critical velocity $v_L$ is determined by the roton part of the spectrum. It is a few times less than  the sound velocity.

The Landau criterion checks the stability with respect to elementary microscopic perturbations of the current state, but does not provide  an information how the instability would develop. The whole process of the supercurrent decay is connected with generation of macroscopic perturbations. These perturbations are quantum vortices. If the vortex axis (vortex line) coincides with the $z$ axis, the phase gradient around the vortex line  is given by 
\begin{eqnarray}
\bm v_v={\hbar\over m}\bm \nabla \varphi_v=\frac{\kappa[\hat z \times \bm r]}{2\pi r^2},
   \label{vort}         \end{eqnarray}
where $\bm r$ is the position vector in the $xy$ plane and $\kappa=h/m$ is the velocity circulation quantum.

\begin{figure}[t]
\includegraphics[width=.3\textwidth]{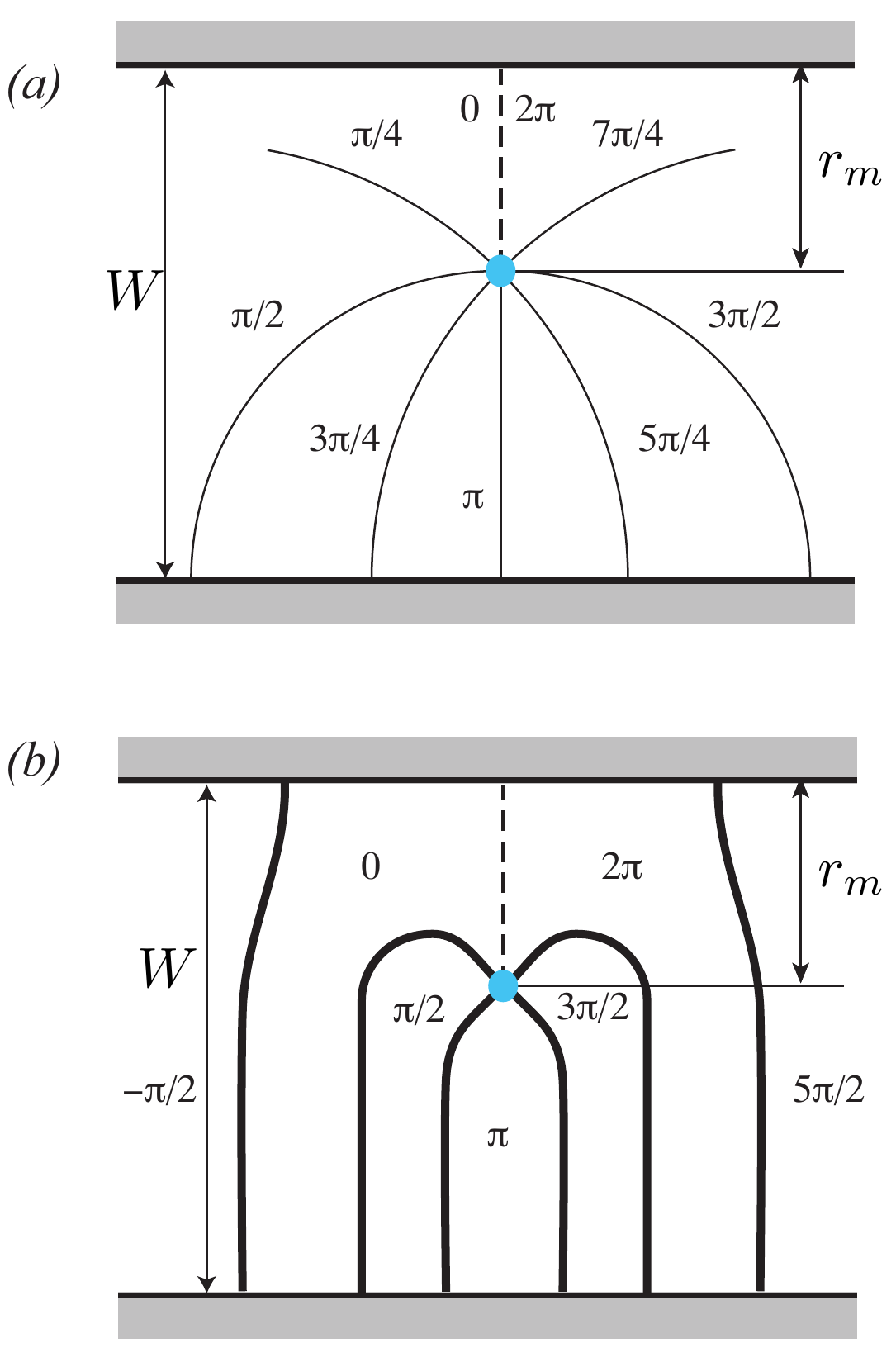}%
\caption{Mass and magnetic  vortices [From \citet{Adv}]. (a) Vortex in a superfluid or magnetic vortex in an easy-plane ferromagnet without in-plane anisotropy. (b) magnetic vortex at small average spin currents ($\langle \nabla \varphi \rangle \ll 1/l$) for four-fold in-plane symmetry. The vortex line is a confluence of four $90^\circ$ domain walls (solid lines).}%
\label{fig3}
\end{figure}

Creation of the vortex requires some energy. The vortex energy per unit length (line tension) is determined mostly  by the kinetic (gradient) energy in the area not very close to the vortex axis where the particle density does not differ essentially from its equilibrium value $n_0$  (the London region): 
\begin{eqnarray}
\varepsilon_v =n_0\int d^2\bm r {\hbar^2 (\bm \nabla \varphi_v)^2\over 2m} = {\pi  \hbar^2 n_0 \over m} \ln{r_m\over r_c}. 
    \label{vorEnM}
          \end{eqnarray}
The upper cut-off $r_m$ of the logarithm is determined by geometry. For the vortex shown in Fig.~\ref{fig3}(a) it is  the distance  of the vortex line from a sample border.
The lower cut-off $r_c$ is the vortex-core radius. It determines the distance $r$ at which the phase gradient is so high that the density $n$ starts to decrease compared to the  equilibrium density $n_0$ at large $r$. The  energy  $E_0(n)$ of the resting superfluid at small $n-n_0$ is determined by the fluid compressibility:
\be
E_0(n)=E_0(n_0)+{m(n_0-n)^2u_s^2\over 2n_0}.
      \ee{}
At the distance $r_c$ from the vortex axis the energy $mn\nabla \varphi_v^2/2$ becomes of the order of the compressibility energy at $n_0-n \sim n_0$.
This happens when the velocity $v_v(r)$ induced by the vortex becomes of the order of the sound velocity $u_s$. This yields $r_c \sim \kappa/u_s$. Inside the core the density vanishes at the vortex axis  eliminating the singularity in the kinetic energy. For the weakly non-ideal Bose-gas $r_c$ is on the  order of  the coherence length.

\begin{figure}[t]
\includegraphics[width=.5\textwidth]{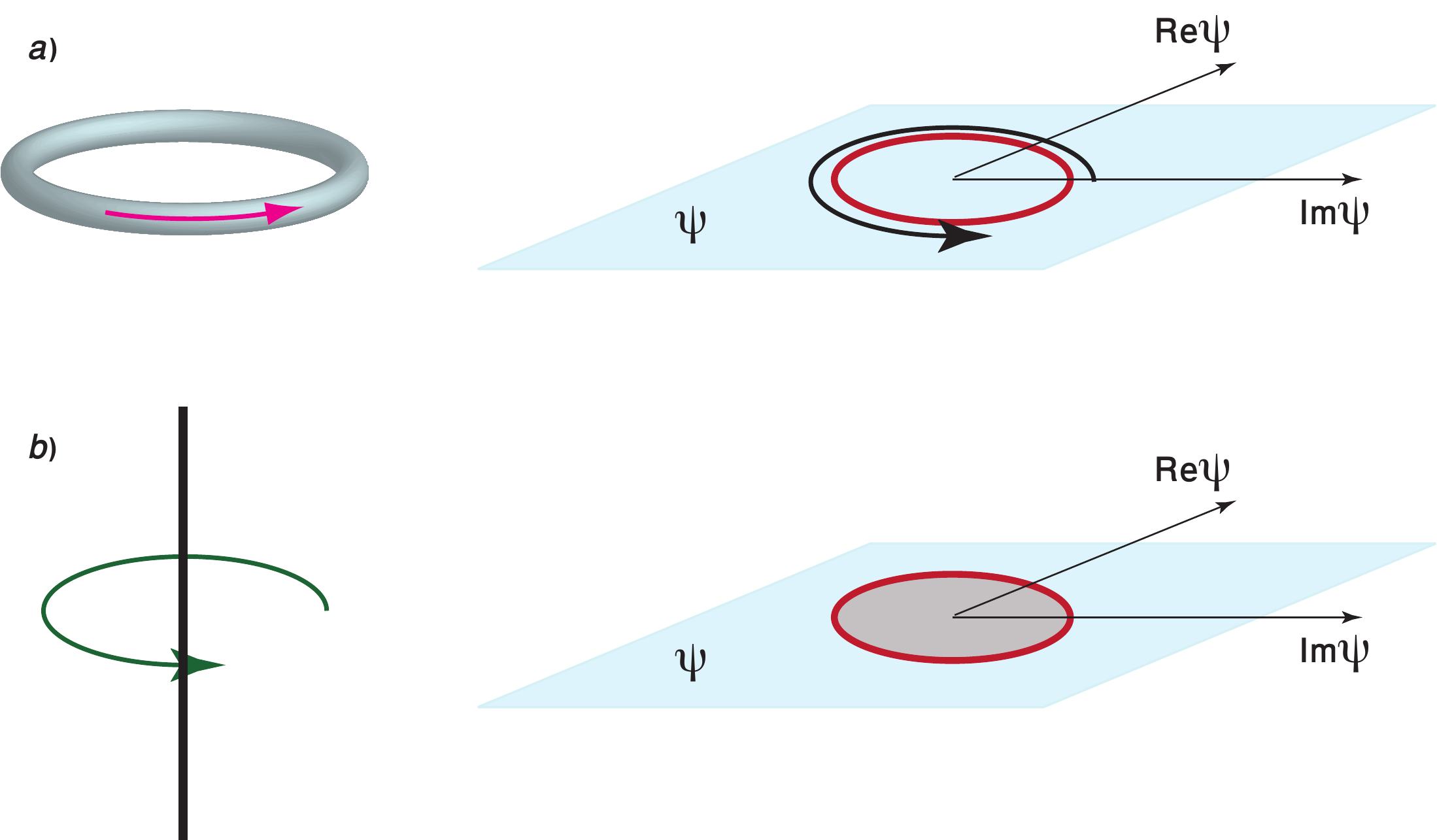}
\caption[]{ Topology of the uniform mass current and the vortex states [From \citet{Adv}]. (a) The current state in a torus maps onto the circumference $|\psi|=\psi_0$ in the complex $\psi$ - plane, where $\psi_0$ is the modulus of the equilibrium order parameter wave function in the uniform state. (b) The  current state with a vortex maps onto the circle $|\psi|\leq \psi_0 $.  }
\label{fig2a}
\end{figure}

Phase slips are impeded  by energy  barriers determined by topology of  the order parameter space (vacuum manifold). The order parameter of a superfluid is a complex wave function   $\psi = \psi_0 e^{i\varphi}$, where the modulus $\psi_0$ of the wave function is a positive constant determined by minimization of the energy and the phase $\varphi$ is a degeneracy parameter.  Any  degenerate ground state   in a closed annular channel (torus)  with some constant phase $\varphi$ maps on some point at the circumference $|\psi|=\psi_0$ in the complex plane $\psi$, while a current state with  the phase change $2\pi N$ around the torus maps onto a path [Fig.~\ref{fig2a}(a)] winding around the  circumference $N$ times. The integer winding number $N$ is a topological charge.  The current can relax if it is possible to change the topological charge.
 
 A  change of the topological charge from $N$ to $N-1$ is possible,  if  a vortex generated at one border of a channel with a moving superfluid moves across current streamlines ``cutting'' the channel  cross-section,  and annihilates at another border as shown in Fig.~\ref{fig3}(a). This is a phase slip. In the phase slip event the distance $r_m$ of the vortex from a border  varies from zero to the width $W$ of the channel. The energy of the vortex in a moving superfluid is determined by a sum of the constant gradient $\bm \nabla \varphi_0$, which determines the supercurrent, and the phase gradient $\bm \nabla \varphi_v$ induced by the vortex. The vortex energy consists of the energy of the vortex in a resting fluid  given by Eq.~(\ref{vorEnM}) and of  the energy from the cross terms of the two gradient fields $\bm \nabla \varphi_0$ and  $\bm \nabla \varphi_v$:
\begin{eqnarray}
\tilde \varepsilon_v = {\pi  \hbar^2 n_0 \over m} L \ln{r_m\over r_c}-  {2\pi  \hbar^2 n_0 \over m} S \nabla \varphi_0, 
   \label{vorCurEnM}
          \end{eqnarray}
where $L$ is the length of the vortex line and $S$ is the area of the cut, at which the phase jumps by $2\pi$. For the 2D case shown in Fig.~\ref{fig3}(a) (a straight vortex in a slab of thickness $L$ normal to the picture plane) $S=Lr_m$. The vortex motion across the channel (growth of $r_m$) is impeded by the barrier determined by the maximum of the energy $\tilde \varepsilon_v$ as a function  $r_m$.  The height of the barrier is the vortex energy at $r_m =1/ 2 \nabla \varphi_0$: 
\begin{eqnarray}
 \varepsilon_b \approx {\pi  \hbar^2 n \over m} L \ln{2\over  r_c \nabla \varphi_0}.
   \label{barrM} 
           \end{eqnarray}
The barrier disappears at gradients $\nabla \varphi_0 \sim 1/r_c$, which are of the same order as the critical gradient determined from the  Landau criterion. On the other hand, in the limit of small velocity $v\propto \nabla \varphi_0$ the barrier height grows and at very small velocity $v\sim \hbar/mW$   reaches the value 
\begin{eqnarray}
 \varepsilon_m \approx {\pi  \hbar^2 n \over m} L \ln{W\over  r_c \nabla \varphi_0}.
   \label{barMM} 
           \end{eqnarray}
 Thus, in the thermodynamic  (macroscopic) limit $W\to \infty$ the barrier height becomes infinite. Since the phase slip probability exponentially decreases on the barrier height (whether the barrier is overcome due to thermal fluctuations or via quantum tunneling) the life time of the current state  in conventional superfluidity diverges when the velocity (phase gradient)  decreases.  This justifies calling superfluidity macroscopic quantum phenomenon.

In the 3D geometry  the phase slip is realized  with expansion of vortex rings. For the ring of radius $R$ the vortex-length and the area of the cut are $L=2\pi R$ and $S=\pi R^2$ respectively, and the barrier disappears at the same critical gradient $\sim 1/r_c$ as in the 2D case.

 The state with the vortex in  a moving superfluid maps  on the full circle $|\psi| \leq \psi_0$ [Fig.~\ref{fig2a}(b)]. The area outside the vortex core maps on the circumference $|\psi|=\psi_0$, while the core maps on the interior of the circle.
 In a weakly interacting Bose-gas the structure of the core is determined by solution of the Gross—Pitaevskii equation \cite{PS,EBS}. The details of the core structure are not important for the content of the present article.

\begin{figure}[b]
\includegraphics[width=.4\textwidth]{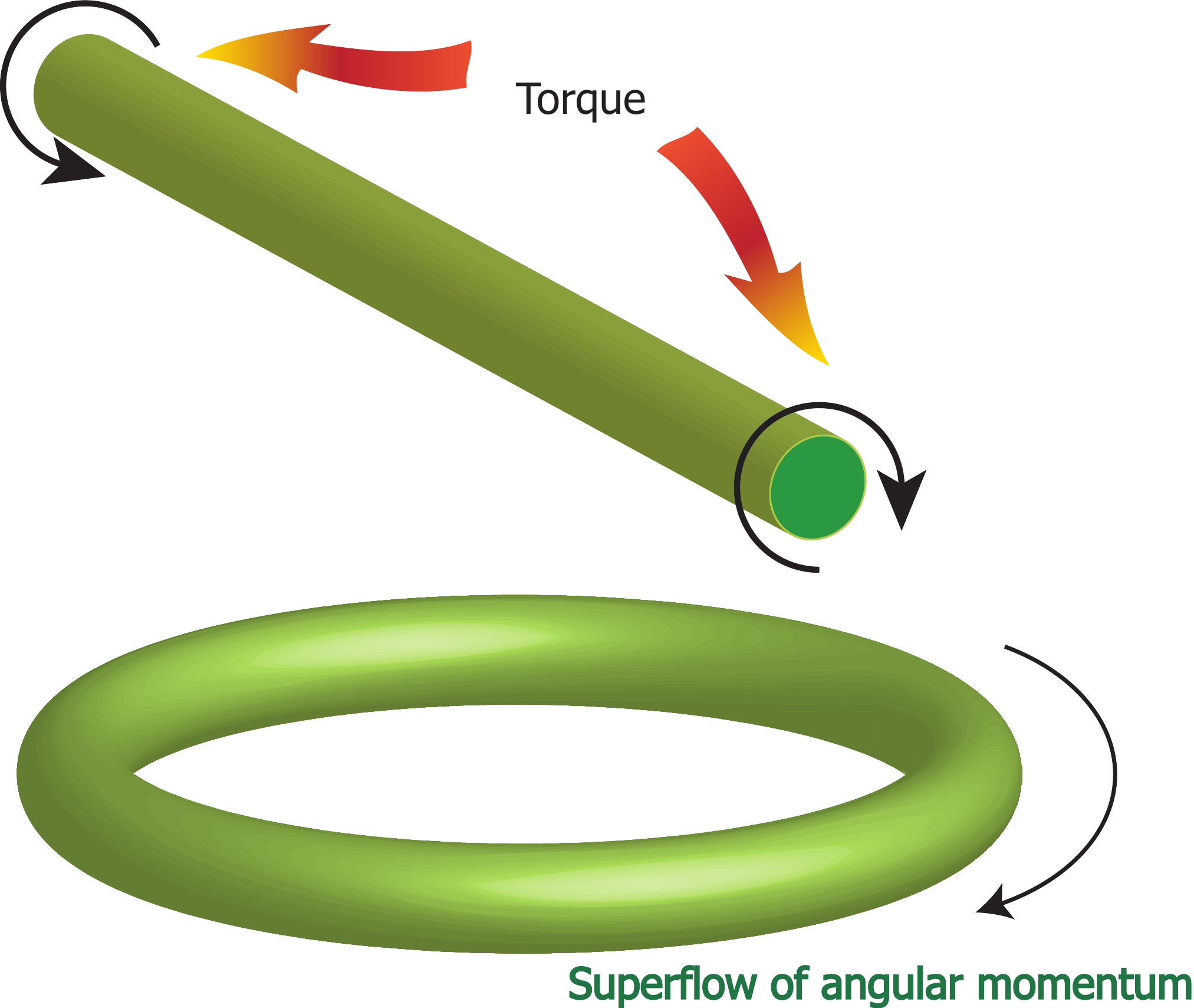}
\caption{Mechanical analogue of a persistent current: A twisted elastic rod bent into a closed ring. There is a persistent angular-momentum flux around the ring [From \citet{Adv}].}%
\label{fig3a}
\end{figure}

For better understanding of the superfluidity phenomenon it is useful to  consider a mechanical analogue of superfluid current \cite{ES-82}. Let  us twist a long elastic rod so  that  a twisting angle at one end of the rod with respect to an opposite end  reaches values many times $2\pi$. Bending the rod into a ring and connecting the ends rigidly, one obtains a ring with a circulating persistent angular-momentum flux (Fig.~\ref{fig3a}). The  flux is proportional to the gradient of twisting angle, which plays the role of the phase gradient in the  supercurrent. The deformed state of the ring is not the ground state of the ring, but it cannot relax to the ground state via any elastic process, because it is topologically stable.  The only way to relieve the strain inside the rod is {\em plastic displacements}. This means that dislocations start to move across rod cross-sections. The role of dislocations in the twisted rod is similar to  the role of vortices in the superfluid.

\section{Spin superfluidity in ferromagnets}\label{magn}

 For a ferromagnet with magnetization density $\bm M$ the LLG equation is \cite{LLstPh2}
\begin{eqnarray}
{\partial \bm M\over \partial t}=\gamma \left[\bm H _{eff} \times  \bm M\right],
     \label{LLP}      \end{eqnarray}
where $\gamma$ is the gyromagnetic ratio between the magnetic and mechanical moment. The effective magnetic field is determined by the functional derivative of the Hamiltonian $\cal H$:
\begin{eqnarray}
\bm H _{eff} =- {\delta {\cal H }\over \delta \bm M}=- {\partial {\cal H }\over \partial \bm M}+\nabla_i{\partial {\cal H }\over \partial \nabla_i\bm M}.
           \end{eqnarray}
According to the   LLG equation, the absolute value $M$ of the magnetization does not vary.  The evolution of  $\bm M$  is a precession around the effective magnetic field $\bm H _{eff}$.        

If spin-rotational invariance is broken and  there is uniaxial crystal magnetic anisotropy the phenomenological Hamiltonian is
\begin{eqnarray}
{\cal H}= {A\over 2} \nabla_i \bm M \cdot \nabla_i \bm M+{G M_z^2\over 2M^2} - \bm H \cdot \bm M.
           \end{eqnarray}
The parameter  $A$ is stiffness of the spin system determined by exchange interaction. If the anisotropy parameter $G$ is positive, it is the ``easy plane'' anisotropy, which keeps the magnetization  in the $xy$ plane. If the external magnetic field $\bm H$ is directed along the $z$ axis, the $z$ component of spin is conserved because of  invariance with respect to rotations around the $z$ axis. For the sake of simplicity we ignore the magnetostatic energy,  which depends on sample shape.

Since the absolute value $M$ of magnetization is fixed, the magnetization vector $\bm M$  is fully determined by the $z$ magnetization component $M_z$ and the angle $\varphi $ showing the direction of $\bm M$ in the easy plane $xy$:
\be
M_x =M_\perp \cos \varphi,~~M_y =M_\perp \sin \varphi,~~M_\perp=\sqrt{M^2-M_z^2} .
    \ee{}
In the new variables the Hamiltonian is
\begin{eqnarray}
{\cal H}= {AM_\perp^2(\bm \nabla \varphi)^2\over 2}+{ M_z^2\over 2\chi}- HM_z  .
  \label{Ener}    \end{eqnarray}
Here we neglected gradients of $M_z$. The magnetic susceptibility $\chi= M^2/G$ along the $z$ axis is determined by the easy-plane  anisotropy parameter $G$.  The LLG equation reduces to the Hamilton equations for a pair of canonically conjugate continuous variables ``angle--moment'':
   \begin{eqnarray}
{1\over \gamma}{d\varphi \over dt}=- {\delta {\cal H} \over \delta M_z}=- {\partial {\cal H} \over \partial M_z},
     \label{HEp} \end{eqnarray}
     \begin{eqnarray}
{1\over \gamma}{dM_z \over dt}={\delta {\cal H}\over \delta \varphi}=-\bm \nabla \cdot{\partial {\cal H} \over \partial \bm \nabla \varphi},
 \label{HEm}      \end{eqnarray}
where functional derivatives are taken from the Hamiltonian Eq.~(\ref{Ener}). Using the expressions for functional derivatives the Hamilton equations are
         \begin{eqnarray}
{1\over \gamma}{d\varphi \over dt}=AM_z(\bm \nabla \varphi)^2-{ M_z-\chi  H\over \chi},
     \label{Ep} \end{eqnarray}
     \begin{eqnarray}
{1\over \gamma} {dM_z \over dt}+ \bm \nabla \cdot \bm J_s =0,
 \label{Em}      \end{eqnarray}
where 
\begin{eqnarray}
\bm J_s=-{\partial {\cal H} \over \partial \bm \nabla \varphi} =-A M_\perp^2  \bm \nabla \varphi
   \label{cur}    \end{eqnarray}
is the spin current. Although our  equations are written for  magnetization, but  not for the spin density, $\bm J_s$ is defined as a current of spin with the spin density $M_z/\gamma$.

\begin{figure}[t]
\includegraphics[width=.5\textwidth]{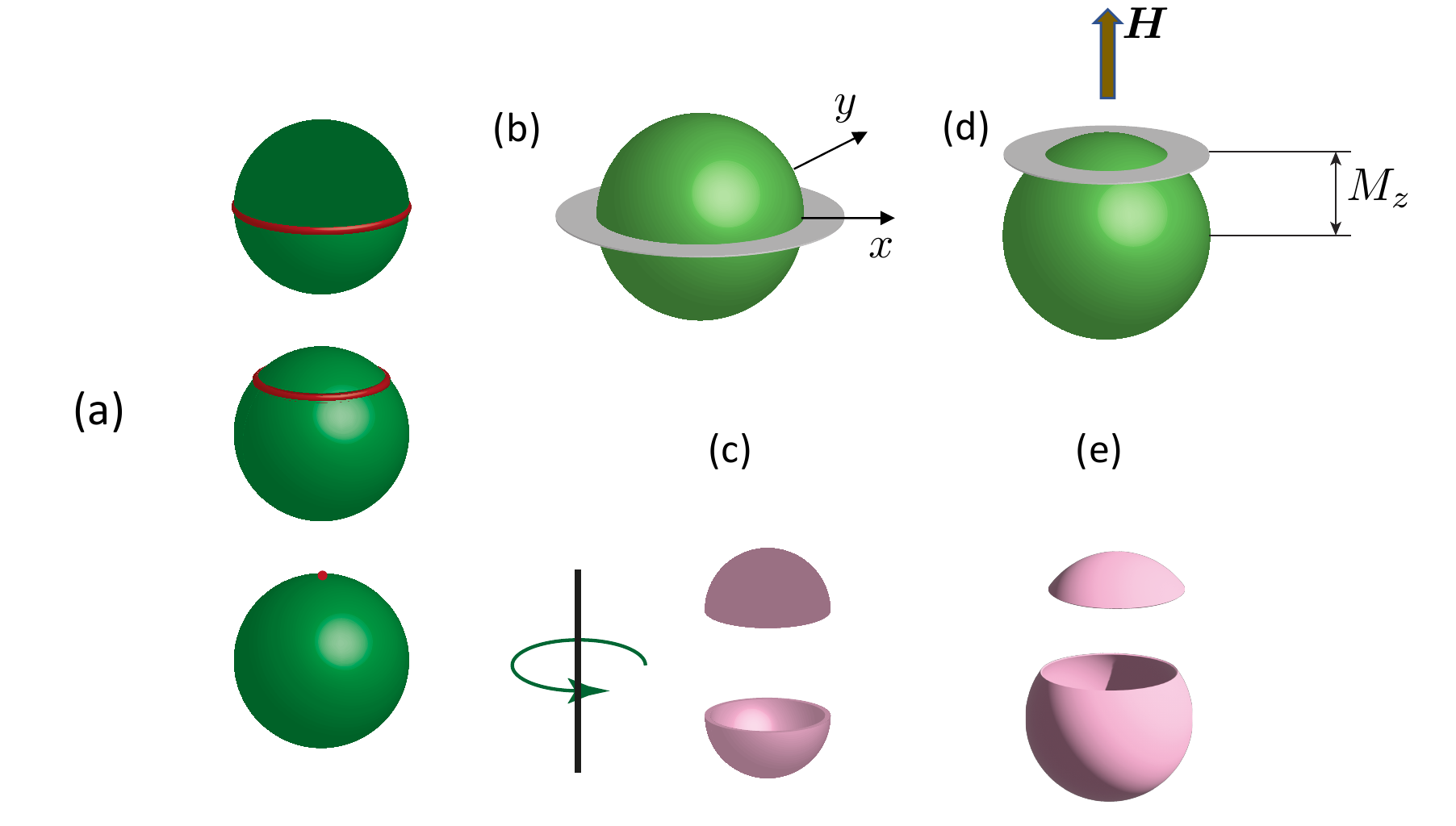}
\caption[]{Mapping of spin current states on the order parameter space of the ferromagnet [From \citet{FNT}].
(a) Spin current in an isotropic ferromagnet. The current state in torus maps on an equatorial circumference on the sphere of radius $M$ (top). Continuous shift of mapping on the surface of the sphere (middle) reduces it to a point at the northern pole (bottom), which corresponds to the ground state without currents. (b) Spin current in an easy-plane ferromagnet at $M_z=0$. The easy-plane anisotropy reduces the order parameter space to an equatorial circumference in the $xy$ plane topologically equivalent to the order parameter space in superfluids. 
(c) Spin current state with a magnetic  vortex  in an easy-plane ferromagnet at $M_z=0$. The states map on the surface of  the upper or the lower  hemisphere. (d) Spin current in an easy-plane ferromagnet at $M_z\neq 0$. Spin is confined in  the plane parallel to the $xy$ plane but shifted  closer to the northern pole.  A  nonzero $M_z $ appears  either  
in the equilibrium due to a magnetic field parallel to the axis $z$, or due to spin pumping. (e)~Spin current state with a magnetic  vortex  in an easy-plane ferromagnet at $M_z\neq 0$. The state maps on the surface of  the upper or the lower  parts of the sphere. Two options of mapping are not degenerate, and the phase slip with the magnetic vortex of a smaller energy (a smaller area of mapping) is more probable.}
\label{Fig02}
\end{figure}

There is an evident analogy of Eqs.~(\ref{Ep}) and (\ref{Em}) with the hydrodynamic equations (\ref{Eul})  and (\ref{IdLiq})  for the superfluid. This analogy of magnetodynamics with 
 hydrodynamics was pointed  out by \citet{HalHoh} for spin waves in antiferromagnets.  The analogy is also important for spin superfluidity.

There are linear solutions of Eqs.~(\ref{Ep}) and (\ref{Em}) describing  the plane spin-wave mode  $\propto e^{i\bm k\cdot \bm r-i\omega t}$ with the sound-like spectrum: 
\be
\omega =  c_{sw}  k,
    \ee{sp0}
where
\be
c_{sw}  =\gamma M_\perp \sqrt{A\over \chi}
    \ee{sv0}
is the spin-wave velocity in the ground state. The variation of the phase in the space is the same as in the  sound mode propagating in the superfluid and shown in Fig.~\ref{fig1}(a).

However, as well as the mass current in a sound wave, the small oscillating spin current in the spin wave does not lead to long-distance superfluid spin transport, which this article addresses. Spin superfluid transport on long distances is realized in current states with spin performing 
 a large number of full 2$\pi$-rotations  in the easy  plane as shown in Fig.~\ref{fig1}(b). In the current state with a constant gradient of the spin phase $\bm K=\bm \nabla \varphi$, there is a constant magnetization component along the magnetic field (the axis $z$):
\be
M_z={\chi H\over 1-\chi AK^2}.
      \ee{Fmag}

 Like in superfluids, the stability of current states is connected with topology of the order parameter space. 
In isotropic ferromagnets ($G=0$) the order parameter space is a spherical surface of radius equal to the absolute value of the magnetization vector $\bm M$ [Fig.~\ref{Fig02}(a)]. All points on this surface correspond  to the same energy of the ground state.  Suppose we created the spin current state with monotonously varying  phase $\varphi$ in a torus. This state maps on the equatorial circumference in the order parameter space. The topology allows to continuously shift the circumference and to reduce it to a point  (the northern or the southern pole).  During this process shown in Fig.~\ref{Fig02}(a)  the path remains in the order parameter space all the time, and therefore, no energetic barrier resists to the transformation. Thus,  in isotropic ferromagnets the metastable current state are not expected.

In a ferromagnet with easy-plane anisotropy ($G>0$) the order parameter space reduces from the spherical surface to a circumference parallel to the $xy$ plane. It is shown in Fig.~\ref{Fig02}(b)  for zero magnetic field when the circumference is the equator. This order parameter space is topologically equivalent to that in superfluids.  Now the transformation of the circumference to the point   costs the anisotropy energy. This allows to expect metastable spin currents (supercurrents). The magnetic field along the anisotropy axis $z$ shifts the easy plane either up [Fig.~\ref{Fig02}(d)] or down away from the equator.

In order to check the Landau criterion, one should know the spectrum of spin waves in the current state with the constant value of the spin phase gradient $\bm K=\bm  \nabla \varphi$ and with the longitudinal (along the magnetic field) magnetization  given by \eq{Fmag}. The spectrum is determined by   solving the Hamilton equations Eqs.~(\ref{Ep}) and  (\ref{Em}) linearized with respect to
weak perturbations of the current state.  We skip the standard algebra given elsewhere \cite{Son19}. Finally one obtains \cite{Hoefer,Son19} the spectrum of plane spin waves:
\be
\omega +\bm w \cdot \bm k = \tilde c_{sw}  k.
    \ee{spF}
Here
\be
\tilde c_{sw}  = \sqrt{ 1 - \chi A K^2} c_{sw} 
      \ee{}
is the spin-wave velocity in the current state, and  the spin wave velocity $c_{sw}$ in the ground state  is given by  \eq{sv0}.
The velocity
\be
\bm w =2\gamma M_z A \bm K
     \ee{}    
can be called Doppler velocity because its effect  on the frequency is similar to the Doppler effect in a Galilean invariant fluid [see the text before \eq{LC}]. However, our system is not Galilean invariant \cite{Hoefer}, and this is the pseudo Doppler effect. Because of it,  the gradient $K$ proportional  to $w$ is present also on the right-hand side of the dispersion relation Eq.~(\ref{spF}). 

We obtained the gapless Goldstone mode with the sound-like  linear in $\bm k$ spectrum.  The current state becomes unstable   when  at $ \bm k$ antiparallel to $\bm w$ the frequency $\omega$ becomes negative. This happens at the gradient  $K$ equal to the Landau critical gradient
\be
K_L={M_\perp \over\sqrt{ 4M^2 - 3M_\perp^2}} {1\over \sqrt{\chi A}}.
     \ee{KcFg}
In the limit of weak magnetic fields when $M_z\ll M$ the Landau critical gradient is
\be
K_L= {1\over \sqrt{\chi A}}= {\gamma  M \over \chi c_{sw} }.
     \ee{KcF}
In this limit the  pseudo-Doppler effect is not important, and  at the Landau critical gradient $K_L$ the spin-wave velocity $\tilde c_{sw} $ in the current state vanishes.

In the opposite limit $M_z \to M$ ($M_\perp \to 0$) the Landau critical gradient, 
\be
K_L={M_\perp \over 2M} {1\over \sqrt{\chi A}},
     \ee{Kcg}
decreases, and the spin superfluidity becomes impossible at the phase transition to the easy-axis anisotropy  ($M_\perp=0$). 

\begin{figure}[b]
\includegraphics[width=.5\textwidth]{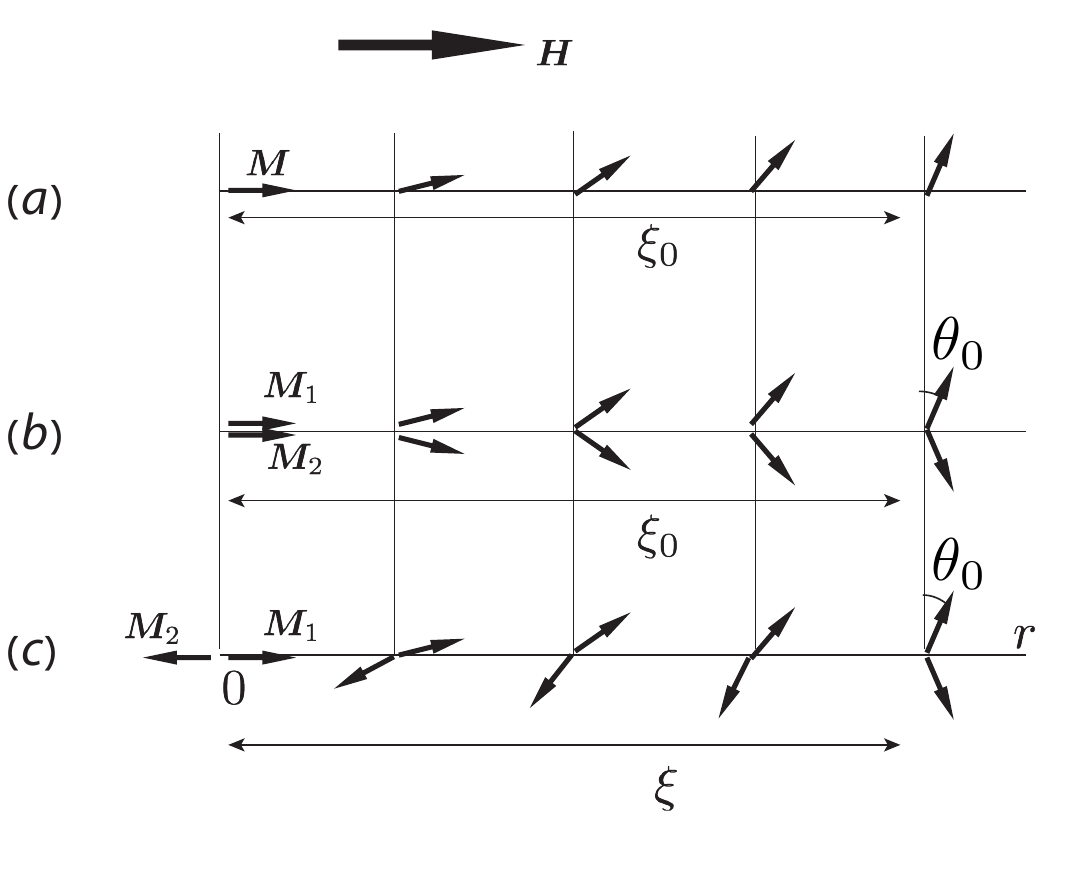}
\caption[]{ Skyrmion cores of magnetic vortices. Variation of magnetization vectors ($\bm M$ in a ferromagnet,  $\bm M_1$ and $\bm M_2$ in an antiferromagnet) in the vortex core as a function of the distance $r$ from the vortex axis is shown schematically. Horizontal directions of magnetizations correspond to the direction of the $z$ axis in the spin space [From \citet{Sp1af}].
(a) The magnetic  vortex in the ferromagnet corresponding to the  gapless Goldstone spin wave mode with the coherence length $\xi_0$ given by \eq{xi0}. 
(b) The magnetic vortex in the antiferromagnet corresponding to the Goldstone spin wave mode with the coherence length  $\xi_0$ given by \eq{Kxi}.   
(c) The magnetic vortex in the antiferromagnet corresponding to the gapped spin wave mode with the coherence length $\xi $ given by \eq{xi}. }
\label{SC}
\end{figure}

Deriving the sound-like spectrum of the spin wave we neglected in the Hamiltonian   terms dependent on gradients  $\bm \nabla M_z$. One should take into account these terms at the wavelength on the order of the coherence length
\be
\xi_0={M\over M_\perp}\sqrt{ \chi A}={M^2\over M_\perp}\sqrt{ A\over G}. 
    \ee{xi0}
The  Landau critical gradient $K_L$ is on the order of the inverse coherence length $1/\xi_0$.

The current states relax to the ground state via phase slips events, in which magnetic vortices cross spin current streamlines.   At $M_z=0$ the current    state with a magnetic  vortex maps on  a surface of a hemisphere of radius $M$ either above or below the equator  \cite{Nik} as shown in Fig.~\ref{Fig02}(c). The vortex core has a structure of a skyrmion. The skyrmion mapping on a  hemisphere is called meron.  At $M_z=0$ two magnetic  vortices with opposite spin polarizations have   the same energy, and both can participate in phase slips. But at $M_z\neq 0$ the magnetic  vortex with a smaller mapping area [Fig.~\ref{Fig02}(e)] has a smaller energy, and phase  slips with its participation are more frequent. Since inside the core the gradients $\bm \nabla M_z$ cannot be ignored, the core radius is on the order  of the coherence length $\xi_0$. Variation of the magnetization direction in space inside the skyrmion core is schematically shown in Fig.~\ref{SC}(a).
One can find more details and numerical calculations of  the structure of skyrmion cores of magnetic  vortices in ferromagnets  elsewhere \cite{Sp1}.

The estimation of barriers for phase slips in spin-superfluid ferromagnets is similar to that  in the case of mass superfluidity. The spin phase gradient in the  current state with a straight magnetic vortex parallel to the axis $z$  is 
\be
\bm \nabla \varphi  ={[ \hat z \times \bm r]\over r^2}+\bm K.
    \ee{phGrad}
We consider a 2D problem of the straight magnetic vortex at the distance $r_m$ from the plane border. The gradient $\bm K$ is parallel to the border.
Substituting the phase gradient field \eq{phGrad} into the kinetic energy and integrating the energy over the London region, where $M_\perp$ is close to its value in the ground state, one obtains the energy of the magnetic vortex per unit length:
\bem
\tilde \varepsilon_v=\pi AM_\perp^2\left(\ln {r_m\over r_c} - 2Kr_m\right). 
   \eem{Venf}
The magnetic vortex energy has a maximum at $r_m =1/2K$. The  energy at the maximum is a barrier preventing phase slips:
\be 
 \varepsilon_b =\pi AM_\perp^2\ln {1\over 2Kr_c}. 
    \ee{Vb}
The barrier vanishes if the gradient $K$ becomes of the order of the inverse vortex core radius $r_c$. This gradient is on the order of the Landau critical gradient $K_L$.

Considering mapping of current states with nonzero magnetization $M_z$ [Figs.~\ref{Fig02}(d) and (e)], we had in mind the equilibrium magnetization $M_z=\chi H$ produced by an external magnetic field $\bm H$. In the equilibrium there is no precession of the magnetization $\bm M$ around the axis $z$. However, non-equilibrium states with non-equilibrium magnetization $M_z$, which makes the magnetization (spin) to coherently precess, are also possible. One can create them  by pumping of magnons, which bring spin and energy into the system. Spin pumping compensates inevitable losses of spin due to spin relaxation. However, usually  the  process of spin relaxation  is weak, and one  may treat  the coherent precession state  as a quasi-equilibrium state at fixed $M_z$. The coherent precession state  does not requires the crystal easy-plane anisotropy for its existence. The easy-plane topology of Fig.~\ref{Fig02}(d) is provided dynamically, and, as a result, metastable spin  current  states are also possible. 

Spin superfluidity  in the quasi-equilibrium coherent precession state  was investigated theoretically and experimentally in the $B$ phase of superfluid $^3$He \cite{Bun}. Later the coherent precession state  in  $^3$He was  rebranded as  magnon BEC \cite{BunV}. Coherent spin precession state (also under the name of magnon BEC) was revealed also
in YIG magnetic films \cite{Dem6}. Spin superfluidity in YIG was discussed by \citet{Pokr,PokrD} and \citet{Son17}.  In the quasi-equilibrium coherent precession state demonstration of the long-distance  superfluid spin transport is problematic (see Sec.~\ref{DC}). Semantic dilemma ``coherent precession state, or magnon BEC'' was discussed by \citet{Adv}. 

\section{Spin superfluidity in antiferromagnets}

The dynamics of a bipartite antiferromagnet can be described by the LLG equations for two spin sublattices coupled via exchange interaction \cite{Kittel51}: 
\be
{d\bm M_i\over dt}=\gamma \left[\bm H_i  \times \bm M_i\right].
       \ee{LLG}
 Here the subscript  $i=1,2$ indicates to which sublattice the magnetization $\bm M_i$ belongs,  and
 \be
\bm H_i =-{\delta {\cal H}\over \delta \bm M_i}= - {\partial {\cal H}\over \partial \bm M_i}+\nabla_j{\partial {\cal H}\over \partial \nabla_j\bm M_i} 
   \ee{}     
is the effective field for  the $i$th  sublattice determined by the functional derivative of the Hamiltonian $\cal H$.  
For an isotropic antiferromagnet  the Hamiltonian is 
\bem
{\cal H}= {\bm M_1 \cdot \bm M_2\over \chi} + { A (\nabla_i \bm M_1 \cdot \nabla_i \bm M_1+\nabla_i \bm M_2 \cdot \nabla_i \bm M_2)\over 2}
\nonumber \\
+A_{12} \nabla_j \bm M_1 \cdot \nabla_j \bm M_2-\bm H \cdot \bm m.~
  \eem{ham2}
The total magnetization is
\be 
\bm m=\bm M_1+\bm M_2, 
 \ee{}
 and the staggered magnetization
 \be
 \bm L=\bm M_1-\bm M_2
   \ee{}
is normal to $\bm m$. 

In the LLG theory absolute values of sublattice magnetizations $\bm M_1$ and $\bm M_2$ are equal to constant $M$, and in the uniform state without gradients  the Hamiltonian is
\be
{\cal H} = -{M^2 \over   \chi}+{m^2 \over  2 \chi} - H  m_z.
  \ee{ham2r}
Here and later on  we assume that the magnetic field is applied along the $z$ axis. The constant term $M^2 /\chi$ can be ignored.  In the ground state the total magnetization $\bm m $ is directed along the magnetic field (the $z$ axis),  and the staggered magnetization $\bm L$ is confined to the $xy$ plane.

The order parameter for an antiferromagnet is the unit N\'eel vector $\bm l=\bm L/L$. The order parameter space for the isotropic antiferromagnet in the absence of the external magnetic field is a surface of a sphere, as for isotropic ferromagnets.  But in  ferromagnets the magnetic field produces an easy axis for the magnetization, while in the antiferromagnet the magnetic field produces the easy plane for the order parameter vector $\bm l$. Thus, the easy-plane topology necessary for the spin superfluidity in antiferromagnets does not require the crystal easy-plane anisotropy.  

In the analogy to the ferromagnetic case, one can describe the vectors of sublattice magnetizations $\bm M_i$ with the constant absolute  value $M$ by the two pairs of the conjugate variables $(M_{iz},\varphi_i)$, which are determined by the two pairs of the Hamilton equations:
 \begin{eqnarray}
{1\over \gamma}{d\varphi_i \over dt}=- {\delta {\cal H} \over \delta M_{iz}}=- {\partial {\cal H} \over \partial M_{iz}},
     \label{aEp} \end{eqnarray}
     \begin{eqnarray}
{1\over \gamma}{dM_{iz} \over dt}={\delta {\cal H}\over \delta \varphi_i}={\partial {\cal H}\over \partial \varphi_i}-\bm \nabla \cdot{\partial {\cal H} \over \partial \bm \nabla \varphi_i}.
 \label{aEm}      \end{eqnarray}

Let us consider the magnetization distribution  with axial symmetry around the axis $z$: 
\bem 
M_{1z}=M_{2z}={m_z\over 2}=M\sin\theta_0,~~m_y=m_z=0,
\nonumber \\
M_{1x}=-M_{2x}={L\over 2}\cos \varphi,
\nonumber \\
M_{1y}=-M_{2y}={L\over 2}\sin \varphi,~~L_z=0.
     \eem{}
Here $\varphi=\varphi_1=\pi-\varphi_2$ is the angle  of rotation of $\bm L$ around the $z$ axis, $L=\sqrt{4M^2-m_z^2}=2M\cos\theta_0$,
and $\theta_0$ is the canting angle. The Hamiltonian \eq{ham2} for the axisymmetric case becomes the Hamiltonian
\be
{\cal H} = {m_z^2 \over  2 \chi} - H  m_z +{A_- L^2 (\nabla \varphi)^2\over 4}
  \ee{hamU}
for the pair of the canonically conjugate variables $(m_z,\varphi)$. The Hamilton equations  for this pair are
   \begin{eqnarray}
{1\over \gamma}{d\varphi \over dt}={A_-m_z(\bm \nabla \varphi)^2\over 2}-{ m_z-\chi  H\over \chi},
     \label{AFEp} \end{eqnarray}
     \begin{eqnarray}
{1\over \gamma} {dm_z \over dt}+ \bm \nabla \cdot \bm J_s =0.
 \label{AFEm}      \end{eqnarray}
Here $A_-=A-A_{12}$, and
\be
\bm J_s=-{A_- L^2\over 2} \bm \nabla \varphi
      \ee{}
is the superfluid spin current. These equations are identical to Eqs.~(\ref{Ep})-(\ref{cur}) for the ferromagnet after replacing the spontaneous magnetization component $M_z$ by the total magnetization component $m_z$, $A$ by $A_-/2$, and $M_\perp $ by $L$.

In the stationary spin current state there is a constant gradient $\bm K=\bm \nabla\varphi$ of the spin  phase and a constant total magnetization
\be
m_z = {\chi H \over 1-\chi A_- K^2/2}.
   \ee{mzA}

\begin{figure}[t]
\includegraphics[width=.5\textwidth]{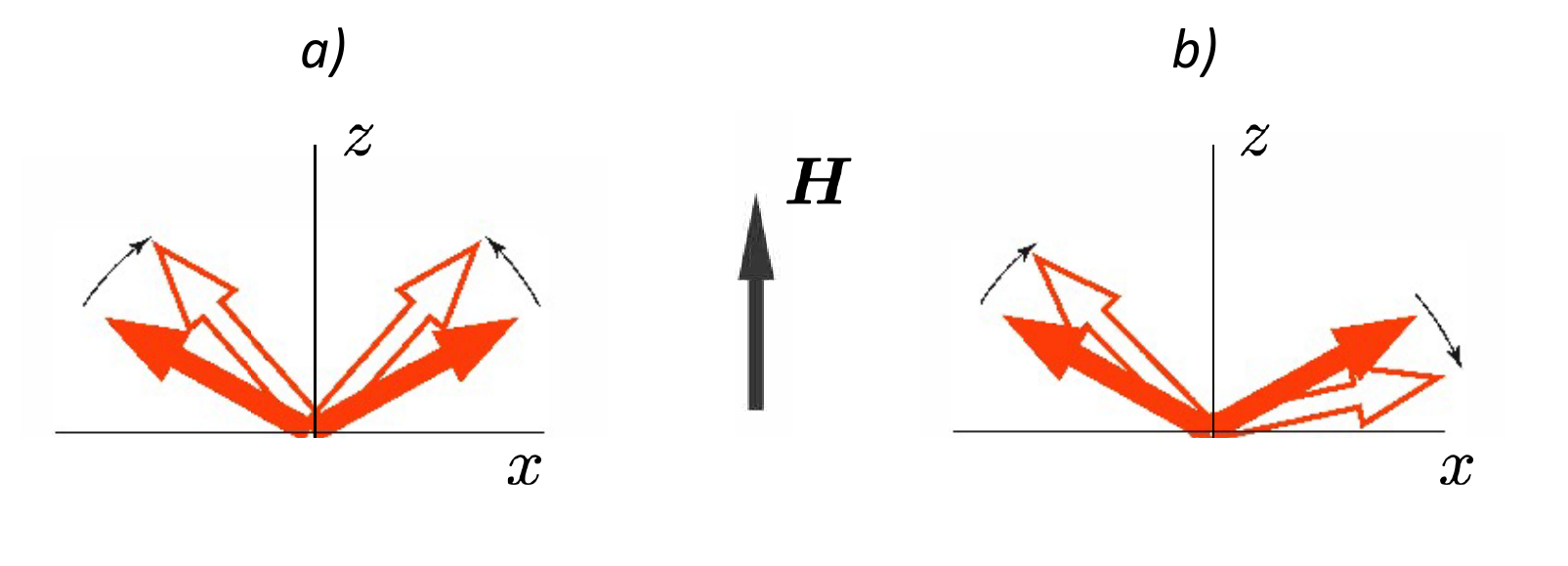}
\caption[]{ The schematic picture of the two spin wave modes in the bipartite antiferromagnet in the plane $xz$  [From \citet{FNT}]. (a) The gapless Goldstone mode. There are oscillations of the canting angle determining the magnetization component $m_z$ and rotational oscillations around the axis $z$.
(b) The gapped mode.  There are rotational  oscillations of the two magnetizations around the axis $y$. }
\label{TM}
\end{figure}

For checking the Landau criterion we must know the spectrum of all collective modes. 
Solution of Eqs.~ (\ref{AFEp}) and  (\ref{AFEm})  linearized with respect  to  plane wave  perturbations $m'\propto  e^{i\bm k\cdot \bm r-i\omega t} $ and $\varphi' \propto  e^{i\bm k\cdot \bm r-i\omega t}$  of the stationary spin current state yield the spectrum of the Goldstone gapless mode:
\bem
\omega +\bm w\cdot \bm k
= \tilde c_{sw} k.
    \eem{}
Here the spin-wave velocity $\tilde c_{sw}$ in the current state and the Doppler velocity $\bm w$ are
\be
\tilde c_{sw}=c_{sw}\sqrt{1 -{\chi A_-  K^2\over 2}},
 ~~ \bm w=\gamma m_z A_-  \bm K,
      \ee{}
while 
\be
c_{sw} =\gamma L \sqrt{ A_-\over 2 \chi}=2\gamma M \cos \theta_0\sqrt{ A_-\over 2 \chi}
     \ee{}
is the spin-wave velocity in the ground  state without spin currents.
The difference between gapless modes in a ferromagnet and an antiferromagnet is that in the former the total magnetization precesses around the $z$ axis, while in the latter there is the precession of the staggered magnetization. Oscillations of sublattice magnetizations $\bm M_1$ and $\bm M_2$ in the gapless mode are illustrated in Fig.~\ref{TM}(a).

In antiferromagnets there is another mode in which the total magnetization  $\bm m$ and the staggered magnetization $\bm L$ perform rotational oscillations around the axis normal to the axis $z$. Without spin supercurrents this axis does not vary in space, and one can choose it to be the axis $y$. The oscillations of the sublattice magnetizations are illustrated in Fig.~\ref{TM}(b).  In the spin current state  $\bm L$ rotates 
around the axis, which itself rotates in the easy-plane $xy$ along the current streamlines. The magnetic field breaks invariance with respect to rotations around axes normal to the field, and the mode spectrum is
\bem
\omega + \bm w\cdot \bm k=\sqrt{\omega_0^2 +c_{sw}k^2} ,
    \eem{spG}
where the gap is given by
\be
\omega_0=\sqrt{ {\gamma^2m_z ^2\over \chi^2}  - c_{sw}^2K^2}.
    \ee{gap}
More details of the derivation are given by \citet{Son19}.

Applying the Landau criterion to the gapless mode at small canting angles $\theta_0$ (weak magnetic fields), one obtains  the critical gradient $K_L$ and the correlation length $\xi_0$,
\be
K_L ={1\over \xi_0},~~      \xi_0  =\sqrt{\chi A_- \over 2},
   \ee{Kxi}  
similar to those obtained for a ferromagnet  [Eqs.~(\ref{Kcg}) and (\ref{xi0})]. However, in contrast to a ferromagnet where the susceptibility $\chi$ is connected with weak anisotropy energy, in an antiferromagnet the susceptibility $\chi$ is determined by a much larger exchange energy and is rather small. As a result, in the antiferromagnet 
the gapped mode loses its stability at much  lower values of $K$ than the gapless mode.  This happens at the Landau critical gradient 
\be
K_L ={1\over \xi}
   \ee{} 
when  the gap given by \eq{gap}  vanishes and the mode frequency becomes complex. Here we introduced a new correlation length
\be
\xi ={c_s\over \gamma H}={\chi c_s \over \gamma m_z}.
    \ee{xi}

As usual, the instability with respect to phase slips with magnetic vortices starts at the gradients  of the same order as the Landau critical gradient. Two modes 
in antiferromagnets have  different correlation lengths, and, correspondingly, there are two types of magnetic vortices  with different structure and size of the skyrmion  core. Figure~\ref{SC}(b) shows schematically spatial variation of the sublattice magnetizations in the skyrmion core of a magnetic  vortex connected with the Goldstone gapless mode. The  core radius is of the order of the correlation length $\xi_0$ given by \eq{Kxi}.
Inside  the core the  canting angle $\theta_0$ grows and reaches $\pi/2$ at the vortex axis.  This transforms an antiferromagnet to a ferromagnet with the magnetization $2M$. The  transformation increases the exchange energy, and at weak magnetic fields  creation of magnetic vortices connected with the gapped mode starts earlier. Its core size  is determined by the larger correlation length $\xi$ determined by the Zeeman energy and given by \eq{xi}. The skyrmion core connected with the gapped mode is  illustrated in Fig.~\ref{SC}(c).

\section{Superfluid  spin transport without spin conservation law}    \label{Phase}

Though processes violating the spin conservation law  are relativistically weak, their  effect is of principal importance and cannot be ignored in general. 
Here we consider the effect of broken rotational symmetry in the easy plane in ferromagnets. Its extension  on antiferromagnets requires insignificant modifications.  
One should  add the $s$-fold in-plane anisotropy energy $\propto G_{in}$ to the Hamiltonian (\ref{Ener}), which  becomes
\be
{\cal H}={ M_z^2\over 2\chi}-\gamma M_z H+ {AM_\perp^2(\bm \nabla \varphi)^2\over 2}+G_{in}[1-\cos (s\varphi)] .
    \ee{an} 
Then the spin continuity equation (\ref{Em}) becomes
   \be
{dM_z \over dt}=-\bm \nabla \cdot \bm J_s+sG_{in} \sin(s\varphi)= AM_\perp^2\left[\nabla^2 \varphi -{\sin(s\varphi)\over l^2}\right] ,
 \ee{EmK}  
where 
 \begin{eqnarray}
l=\sqrt{ AM_\perp^2\over sG_{in}}
             \end{eqnarray}
is the thickness of the wall separating domains with  $s$ equivalent easiest directions in the easy plane. 
In the stationary spin current states $dM_z/dt=0$, and the  phase $\varphi$ is a periodical solution of the sine-Gordon equation parametrized by the average phase gradients $\langle \nabla \varphi \rangle$. At small $\langle \nabla \varphi \rangle \ll 1/l$ the spin structure constitutes the chain of domains with the period $2\pi/s\langle \nabla \varphi \rangle$. Spin currents (gradients) inside domains are negligible but there are spin currents inside domain walls.
The spin current has a maximum in the center of the domain wall  equal to
\be
J_l=\sqrt{2\over s}{A\over l}. 
     \ee{curL}

\begin{figure}[b]
\includegraphics[width=.35\textwidth]{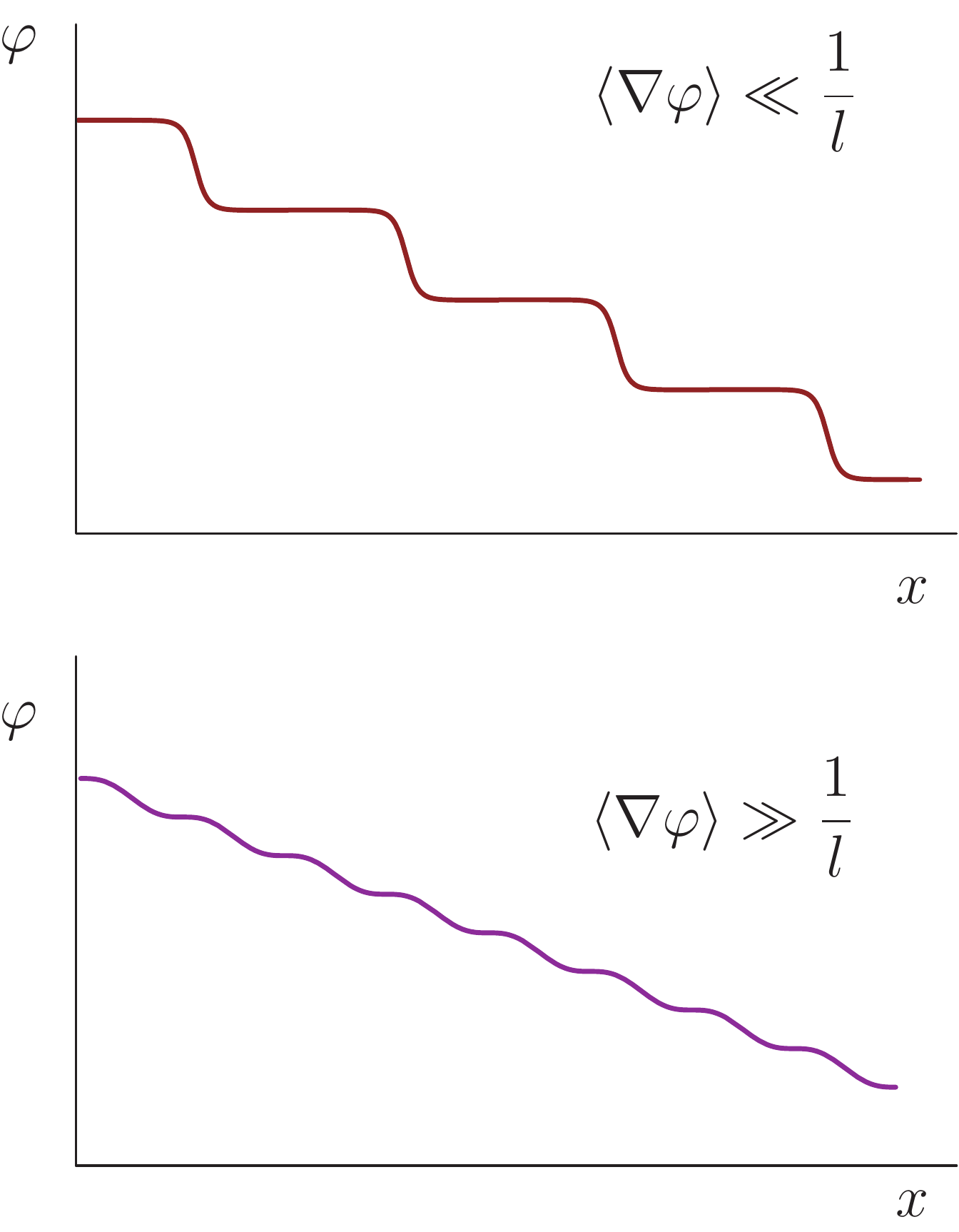}
\caption[]{ The nonuniform spin-current states with $\langle \nabla \varphi \rangle \ll 1/l$ and $\langle \nabla \varphi \rangle \gg 1/l$ [From \citet{Adv}]. }
\label{fig2}
\end{figure}

The spin transport in the case $\langle \nabla \varphi \rangle \ll 1/l$ hardly reminds the  superfluid transport on macroscopic scales: spin is transported over distances on the order of the domain-wall thickness $l$. With increasing $\langle \nabla \varphi \rangle$ the density of domain walls grows, and at $\langle \nabla \varphi \rangle\gg 1/l$ the domains coalesce. Deviations of the gradient $\nabla \varphi$ from the constant average gradient $\langle \nabla \varphi\rangle$  become negligible. This restores the analogy with the superfluid transport in superfluids \cite{ES-78a,ES-78b,ES-82}, and  spin non-conservation can be ignored. The transformation of the domain wall chain into a weakly inhomogeneous current state at growing $\langle \nabla \varphi \rangle$  is illustrated in Fig.~\ref{fig2}. According to the Landau criterion, spin current states become unstable at $\nabla \varphi 
\sim 1/\xi_0$, where the correlation length $\xi_0$ is given by  \eq{xi0}. Thus, one can expect metastable nearly uniform current states at $1/l \ll \langle \nabla \varphi \rangle \ll 1/\xi_0$. This is possible if the easy plane anisotropy energy $G$ essentially exceeds the in-plane anisotropy energy $G_{in}$.

In the limit $\nabla\varphi \ll 1/l$ of strongly nonuniform current states the decay of the current is also possible only via phase slips, but the structure of magnetic vortices is essentially different from that in the opposite limit $\nabla\varphi \gg 1/l$. The magnetic vortex is a line defect at which $s$ domain walls end. The phase slip with the magnetic  vortex crossing the streamlines in the channel leads to annihilation of $s$ domain walls. This process is illustrated in Fig.~\ref{fig3}(b) for the four-fold in-plane symmetry ($s=4$).

An important difference with conventional mass superfluidity is that while the existence of conventional superfluidity is restricted only from above by the Landau critical gradients,  the spin superfluidity is restricted also from below: gradients should not  be less than the value $1/l$. Thus, the gradient $K=\langle \nabla \varphi \rangle$ in the expression \eq{Vb} barrier height cannot be less than $1/l$, and the height of the barrier cannot exceed 
\be 
 \varepsilon_m =\pi AM_\perp^2\ln {l\over r_c}. 
    \ee{barS}
In contrast to the maximal phase slip barrier \eq{barMM} in mass superfluidity, in spin superfluidity the maximal phase slip barrier does not  become infinite in the macroscopic limit $W\to \infty$ \cite{ES-78a,ES-78b,ES-82,McD}.

\section{Long-distance superfluid spin transport } \label{SpTr}

\begin{figure}[b]
\includegraphics[width=.4\textwidth]{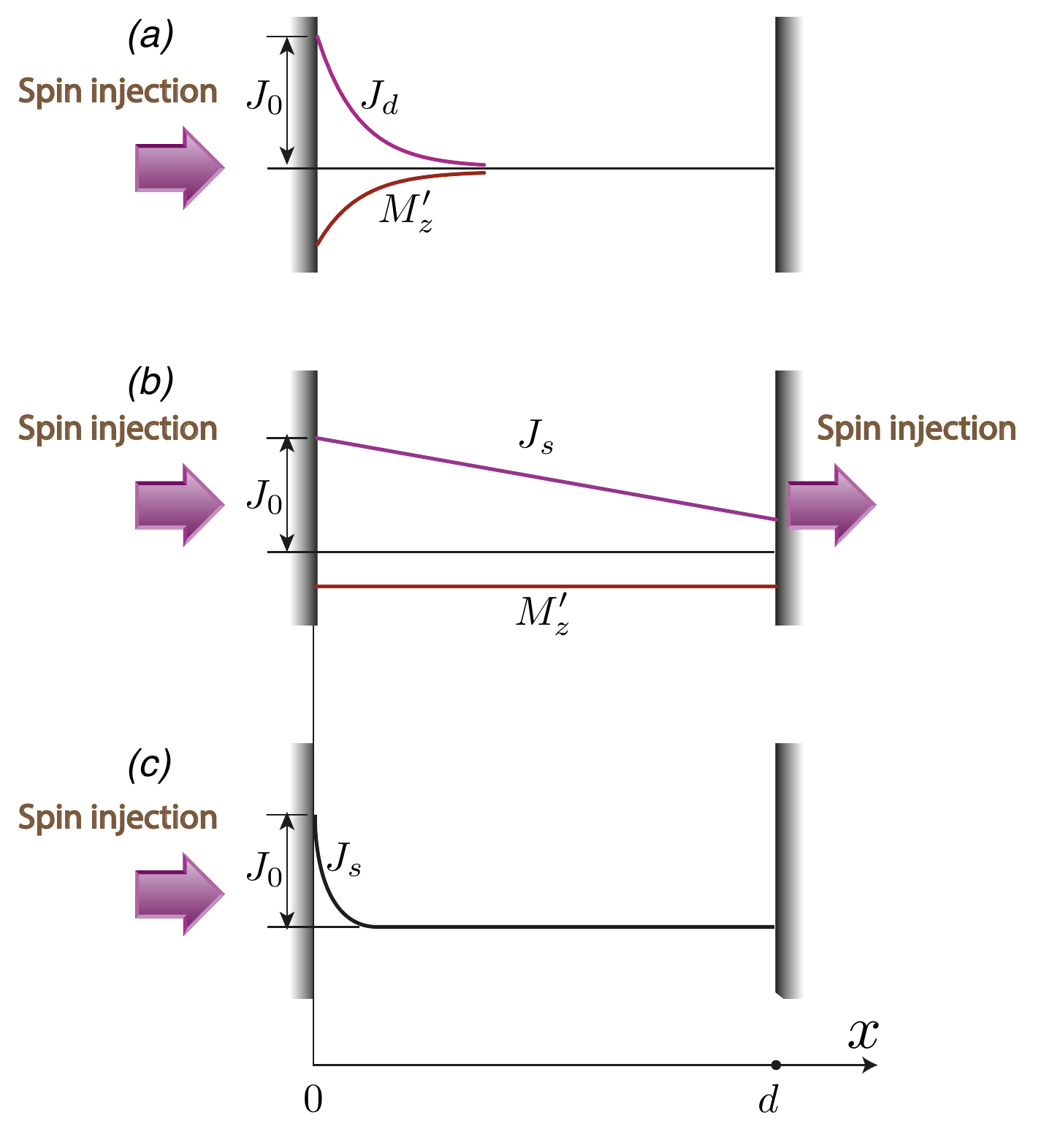}
\caption[]{ Long distance spin transport [From \citet{Adv}]. (a) Injection of the spin current $J_0$ into a spin-non-superfluid medium. (b) Injection of the strong spin current $J_0 \gg J_l$ into a  spin-superfluid medium. (c) Injection of the weak spin current $J_0 <J_l$ into a spin-superfluid medium.  }
\label{f1}
\end{figure}

The absence of the strict spin conservation law also leads to the dissipative process, which is impossible in the  mass superfluidity and very important for long-distance superfluid spin transport \cite{ES-78b,Adv,Tserk,Halp}: longitudinal spin relaxation characterized in the magnetism theory by the Bloch time $T_1$. Taking the Bloch relaxation into account, the equation for the non-equilibrium magnetization $M'_z=M_z-\chi H$ becomes
    \be
{1\over \gamma} {dM'_z \over dt}=-\bm \nabla \cdot \bm { J}- {M'_z \over \gamma T_1}. 
      \ee{EmB}
Here the total current $\bm { J}=\bm J_s  +\bm J_d$ includes not only the spin supercurrent $\bm J_s$ given by \eq{cur}, but also the spin diffusion current 
    \be
 \bm  J_d =- {D\over \gamma}\bm\nabla M_z.
       \ee{Jd}

 In the absence of spin superfluidity ($\bm J_s=0$)  \eq{EmB} describes pure spin diffusion [Fig.~\ref{f1}(a)].  Its solution, with the boundary condition that the spin current $J_0$ is injected at  the interface $x=0$, is
\be
J_d=J_0e^{-x/L_d}, ~~M'_z=\gamma J_0\sqrt{T_1\over D}  e^{-x/L_d},
   \ee{}
where 
\be
L_d=\sqrt{DT_1} 
   \ee{}
is the spin-diffusion length. Thus, the effect of spin injection exponentially decays at the scale of the spin-diffusion length,  and the density of  spin accumulated at the other border of the medium decreases exponentially  with growing distance $d$.

However, if spin superfluidity is possible, the spin precession equation Eq.~(\ref{Ep}) becomes relevant. According to this equation, in a stationary state the magnetization $ M'_z$ cannot vary in space [Fig.~\ref{f1}(b)]  since the gradient $\bm\nabla M'_z$ leads to the linear in time growth of the gradient $\bm\nabla \varphi$. 
The right-hand side of \eq{Ep} is an analog of the chemical potential, and the requirement of constant in space magnetization $M_z$  is similar to the requirement of constant in space chemical potential in superfluids, or the electrochemical potential in superconductors.  As a consequence of this requirement, spin diffusion current is impossible in the bulk since  it is simply ``short-circuited'' by the superfluid spin current. The bulk spin diffusion current can appear only in AC processes.

If the spin superfluidity is possible, the spin current can reach the spin detector at the plane $x=d$  opposite to the border where spin is injected.  As a boundary condition at $x=d$, one can use a phenomenological relation $J_s(d) = M'_z(d) v_d$ connecting the spin current with the non-equilibrium magnetization at the border with a non-magnetic medium. Here $v_d$ is a phenomenological constant. This boundary condition \cite{ES-78b} was confirmed by the microscopic theory of \citet{Tserk}. Together with the boundary condition $J_s(0) = J_0$ at $x=0$ this yields the solution of Eqs.~(\ref{Ep}) and (\ref{EmB}):
 \be
M'_z= { T_1 \over d+v_dT_1}\gamma J_0,~~J_s(x) = J_0 \left(1-{x\over d+v_d  T_1} \right).
     \ee{LDtr}
Thus, the spin accumulated at large distance $d$ from the spin injector slowly decreases with $d$ as  $1/(d+C)$ [Fig.~\ref{f1}(b)], in contrast to the exponential decay $\propto e^{-d/L_d}$ in the spin diffusion transport [Fig.~\ref{f1}(a)]. The constant $C$ is determined by the boundary condition at $x=d$.

The long-distance superfluid spin transport is possible only if the injected spin current is not too small. If  the injection spin current $J_0$ is less than the current $J_l$ determined by \eq{curL} the superfluid  spin current  penetrates into the medium  only at distances not longer than the width $l$ of a domain wall in a non-uniform spin current state at a very small average gradient $\langle \nabla \varphi \rangle \ll 1/l$ [Fig.~\ref{f1}(c)]. This  threshold for  the long-distance spin superfluid transport is connected with the absence of the strict conservation law for spin (Sec.~\ref{Phase}).

\section{Experiments on detection of spin superfluidity} \label{DC}

Experimental detection of spin superfluidity  does not reduce to experimental evidence of  the existence of spin supercurrents proportional to gradients of spin phase.  As pointed out in Introduction (Sec.~\ref{Intr}), such supercurrents produced by spin phase difference smaller than $2\pi$ emerge in any non-uniform spin structure. 
Numerous  observations of spin waves and domain structures during the more than half-a-century history of modern magnetism
 cannot be explained without these microscopic spin currents. Only detection of macroscopical spin supercurrents produced by the phase difference many times larger than $2\pi$  is evidence of spin superfluidity. 

The experimental evidence  of macroscopical spin supercurrents  was obtained in the past in the {\it B} phase of superfluid $^3$He \cite{flux}.   A spin current was generated in a long channel connecting two cells filled by $^3$He-{\it B}.  The quasi-equilibrium state of the coherent spin precession  was supported by spin pumping. The magnetic fields applied  to the two cells were slightly different, and therefore, the spins in the two cells precessed  with different frequencies. A small difference in the frequencies  leads to a linear growth of
difference of the precession phases in the cells and a phase gradient in the channel. When the gradient reached the critical value, $2\pi$ phase slips were detected. 
This was evidence of non-trivial spin supercurrents. 

This experiment was done in the dynamical state of coherent spin precession (non-equilibrium magnon BEC). The  states require pumping of spin in the whole bulk for their existence. In the geometry of the experiment  on  long-distance  spin transport (Sec.~\ref{SpTr}) this would  mean that spin is permanently pumped not only by a distant injector but also all the way up to the place where its accumulation is probed. Thus, the spin detector measures  not only spin coming from the distant injector but also  spin pumped close to the detector. Therefore, the experiment cannot demonstrate the existence of long-distance superfluid spin  transport, but can provide, nevertheless, indirect evidence that long-distance superfluid spin  transport is possible in principle. 
 
\begin{figure}[t]
\includegraphics[width=.4\textwidth]{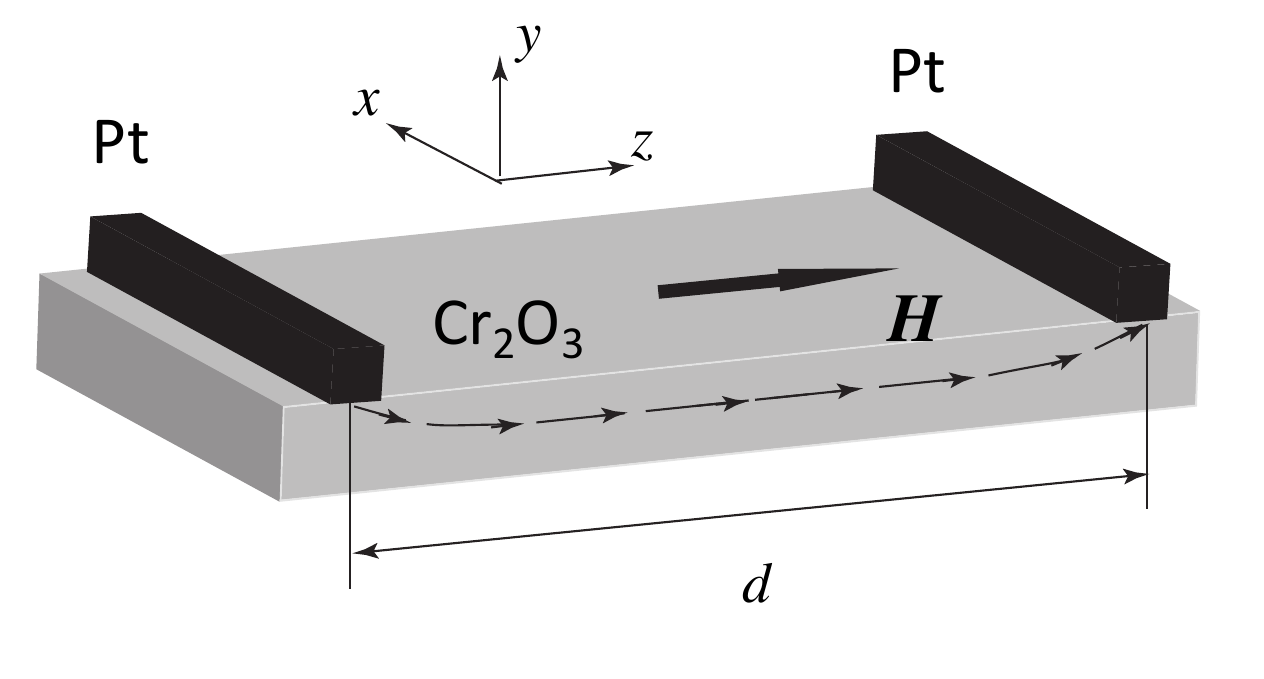}
\caption[]{ Long distance spin transport in the experiment by \citet{WeiH}.
Spin is  injected from the left Pt wire and flows along the Cr$_2$O$_3$ film  to the right Pt wire, which serves as a detector. The arrowed dashed line shows a spin-current streamline  [From \citet{FNT}]. }
\label{fy}
\end{figure}

The experiment on detection of long-distance superfluid spin transport (Sec.~\ref{SpTr}) was recently done by \citet{WeiH} in antiferromagnetic Cr$_2$O$_3$. The spin was injected from a Pt injector by heating [the Seebeck effect \cite{Seki}] on one side of the Cr$_2$O$_3$ film and spin accumulation was probed on another side of the film by the Pt detector via the inverse spin Hall effect (Fig.~\ref{fy}). In agreement with the theoretical prediction, they observed spin accumulation inversely proportional to the distance from the interface where spin was injected.  

 In the experiment of \citet{WeiH}  spin injection required  heating of the Pt injector,  and the spin current to the detector is inevitably accompanied by a heat flow. \citet{Lebrun} argued that \citet{WeiH} detected a signal not from spin coming from the distant injector but from spin generated by the Seebeck effect at the interface between the heated antiferromagnet and the Pt detector.  If true,  \citet{WeiH} observed not long-distance spin transport but long-distance heat transport. 
A resolution of this controversy requires further experimental and theoretical investigations. In particular, one could check  long-distance heat transport scenario by replacing the Pt spin injector in the experiment of \citet{WeiH} by a heater, which produces  heat but no spin.

Observation of the long-distance  superfluid spin transport was also reported by \citet{Lau18} in a graphene quantum Hall antiferromagnet.  However, the discussion of this report requires an extensive theoretical analysis of the $\nu=0$ quantum Hall state of graphene, which goes beyond the scope of the present article. A reader can find this analysis in \citet{SupGraph}.

\section{Discussion and conclusions}\label{disc}

The article addressed the basics of the spin superfluidity concept: topology, Landau criterion, and phase slips. Metastable (persistent) superfluid current states are possible if the order parameter space (vacuum manifold) has the topology of a circumference like in conventional superfluids. In ferromagnets it is the circumference on the spherical surface of the spontaneous magnetizations $\bm M$, and in antiferromagnets it is the spherical surface of the unit N\'eel vector $\bm L/L$, where  $\bm L$ is the staggered magnetization. The topology necessary for spin superfluidity requires the magnetic easy-plane anisotropy in ferromagnets, while in antiferromagnets this anisotropy is provided by the Zeeman energy, which confines the N\'eel vector in the plane normal to the magnetic field.

The Landau criterion was checked  for the spectrum of elementary excitations, which are spin waves in our case. In ferromagnets there is only one gapless Goldstone spin wave mode. In bipartite antiferromagnets there are two modes: the Goldstone mode in which spins perform rotational oscillations around the symmetry axis and the gapped mode with rotational oscillations around the axis normal to the symmetry axis. At weak magnetic fields the Landau instability starts not in the Goldstone mode, but in the gapped mode.  In contrast to superfluid mass currents in conventional superfluids, metastable spin superfluid currents are restricted not only by the Landau criterion from above but also from below. The restriction from below is related to the absence of the strict conservation law for spin. 

The Landau instability with respect to elementary excitations is a precursor for the instability with respect to phase slips.  The latter instability starts  when the spin phase gradient reaches the value of the inverse vortex core radius. This value is on the same order of magnitude  as the Landau critical gradient.  Magnetic vortices participating in phase slips have skyrmion cores, which map on the upper or lower part of the spherical surface in the space of spontaneous magnetizations in ferromagnets, or in the space of the unit N\'eel vectors  in antiferromagnets. 

It is worthwhile to note that in reality it is not easy to reach  the critical gradients discussed in the present article experimentally. The decay of superfluid spin currents is possible also at subcritical spin phase gradients since the barriers for phase slips can be overcome by thermal activation or macroscopic quantum tunneling.  This makes the very definition of the real critical gradient rather ambiguous  and dependent on duration of observation of persistent currents.   Calculation of real critical gradients requires a detailed dynamical analysis of  processes of thermal activation or macroscopic quantum tunneling through phase slip barriers, which is beyond the scope of the present article. One can find examples of such analysis for conventional superfluids with mass supercurrents in \citet{EBS}.
  
Experimental  evidence of the existence of metastable superfluid spin currents in the $B$ phase of superfluid $^3$He  was reported long ago \cite{flux}. The experiment  was done  in the non-equilibrium state of coherent spin precession, which requires permanent spin pumping in whole bulk  for its existence. This does not allow  to check true long-distance superfluid spin transport without any additional  spin injection on the  way from an injector to a detector of spin. The experiment demonstrating  the long-distance transport of spin in the solid antiferromagnet was reported recently \cite{WeiH}. But the interpretation of this experiment in the terms of spin superfluidity was challenged \cite{Lebrun}, and experimental verification  of the long-distance superfluid spin transport in magnetically ordered solids convincing many (if not all) in the community is still wanted.

Mechanical analogy of the mass superfluidity discussed in the end of Sec.~\ref{MS} is valid also for spin superfluidity.  The  ``superfluid'' flux of the angular momentum in a twisted elastic rod 
is similar to the superfluid spin current in magnetically ordered solids. Of course, it is not obligatory to discuss the twisted rod in terms of angular-momentum flux. More usual is to discuss it in the terms of  the elasticity theory: deformations, stresses, and elastic stiffness. On the same grounds, one can avoid to use the terms  ``spin  current'' and ``spin superfluidity'' and consider the spin current states as metastable helicoidal  spin structures determined by  ``phase stiffness''. This stance was quite popular in early  disputes about spin superfluidity. Nowadays in the era spintronics the terms ``spin supercurrents'' and ``spin superfluidity'' are widely accepted.

In this article the essentials of spin superfluidity were discussed for simpler cases of a ferromagnet and of a bipartite antiferromagnet at    zero temperature. Spin superfluidity was also investigated in antiferromagnets with a more complicated magnetic structure \cite{BoKoval}. At finite temperates the presence of the gas of incoherent magnons  was taken into account in the two-fluid theory \cite{TwoFlu} similar to the two-fluid theory in the theory of mass superfluidity.

The present article focused on spin superfluidity  in  magnetically ordered solids. In superfluid $^3$He spin superfluidity coexists with mass superfluidity. 
Recently investigations of spin superfluidity were extended to spin-1 BEC, where spin and mass superfluidity also
coexist and interplay \cite{LamSpin,Duine,Sp1,Sp1af}.
This interplay leads to a number of new nontrivial features of the phenomenon of superfluidity. The both types of superfluidity are restricted by the Landau criterion for the softer  
collective modes, which usually are the spin wave modes. As a result, the presence of spin superfluidity diminishes the possibility of the conventional mass superfluidity. Another  consequence of the coexistence of spin and mass superfluidity is phase slips with bicirculation vortices characterized by two topological charges (winding numbers) \cite{Sp1af}.


%



\end{document}